\pdfoutput=1
\documentclass[twoside,aps,showpacs,twocolumn]{revtex4}
\usepackage{amsmath,amssymb,graphicx,dsfont}
\usepackage[english]{babel}
\usepackage[utf8]{inputenc}
\usepackage{ stmaryrd }
\usepackage[pdftex]{color}

 \DeclareMathOperator{\erf}{\text{erf}}

\begin{document}

\title{Carnot process with a single particle}
\author{J. Hoppenau}
\email{johannes.hoppenau@uni-oldenburg.de}
\author{M. Niemann}
\email{markus.niemann@uni-oldenburg.de}
\author{A. Engel}
\email{andreas.engel@uni-oldenburg.de}
\affiliation{
Institut f\"{u}r Physik, Carl-von-Ossietzky Universit\"{a}t, 26111
  Oldenburg, Germany }

\begin{abstract}
We determine the statistics of work  in isothermal volume changes of a classical  ideal gas consisting of a single particle. Combining our results with the findings of Lua and Grosberg [J. Chem. Phys. B {\bf 109}, 6805 (2005)] on adiabatic expansions and compressions we then analyze the joint probability distribution of heat and work for a microscopic, non-equilibrium Carnot cycle. In the quasi-static limit we recover Carnot efficiency, however, combined with non-trivial distributions of work and heat. With increasing piston speed the efficiency decreases. The efficiency at maximum power stays within recently derived bounds.
\end{abstract}
\maketitle

\section{Introduction}
\label{sec:intro}

Thermodynamics has to be modified for systems so small that typical changes of their energy are of the order of the thermal energy per degree of freedom. To account for the correspondingly strong fluctuations of energy, work, heat, and entropy, in the emerging field of stochastic thermodynamics thermodynamic quantities are characterized by probability distributions \cite{Udorev,Jarrev,Esprev}. These distributions exhibit unexpected symmetries, now subsumed under the notion of fluctuation theorems, which have revealed new insights into the role of large deviations in statistical mechanics and the foundations of macroscopic thermodynamics. Thanks to recent advances in experimental techniques these findings are amenable to experimental verification and contribute to the understanding of energy conversion on the nano-scale. A particularly intriguing application concerns the efficiency of biological motors \cite{Udobioeta,Chrisbioeta} as well as miniaturized heat engines  \cite{BlBe}. 

One of the remarkable features of fluctuation theorems is that they hold for (almost) arbitrary deviations from thermodynamic equilibrium. In the attempts to intuitively understand this surprising generality case studies of exactly solvable model systems have played a decisive role \cite{JaMa,ZoCo,Lua2005}. Given the ubiquitous presence of the ideal gas model in statistical physics it is not surprising that among these studies those making use of ideal gases \cite{Lua2005,Bena,ClBrKa,Baule2006,NoEn} are particularly prominent. In 2005 Lua and Grosberg \cite{Lua2005} scrutinized the validity of the Jarzynski equality \cite{Jar97} for the adiabatic expansion and compression of a classical ideal gas consisting of one particle and highlighted the importance of the far tails of the Maxwell distribution. Soon afterwards an analogous treatment of isothermal volume changes was pursued \cite{Baule2006}, however, analytical progress was limited to the approximate treatment of the case of large piston speed and small volume change. To characterize the general case only numerical simulations were used.  

In the present paper we re-analyze the isothermal expansion and compression of a single classical gas particle in the framework of stochastic thermodynamics and show how some of the obstacles hampering analytical progress can be overcome. Analogous to Lua and Grosberg we consider the particle being inside a cylinder with a moveable piston and make no a-priori assumptions about the piston speed such that also strong deviations from equilibrium are possible. We determine the distribution of work performed by the gas particle as well as the statistics of the heat exchanged with the reservoir. As a test of our findings we check their consistency with the Jarzynski equality and the Crooks fluctuation theorem \cite{Crooks}.

We then consider a cyclic process of Carnot type consisting of two adiabatic and two isothermal strokes. Combining the results of Lua and Grosberg for the adiabatic parts and our own findings characterizing the isothermal changes we determine the joint distribution of work and exchanged heats for one cycle of the engine. We show that even in the quasi-static limit of slow piston speeds fluctuations prevail and that the properties differ from those of the classical Carnot process. Finally, with the distributions of work and heat at hand, we analyze the efficiency of the engine at maximal power and compare the result with previous studies \cite{IzOk1,IzOk2} and with recently derived bounds \cite{EsLiBr,ScSe,EsKaLiBr}.

The paper is organized as follows. In section \ref{sec:model} we introduce our model and fix the notation. Section \ref{sec:isoth} contains the analysis of an isothermal expansion and compression and parallels the treatment of Lua and Grosberg for adiabatic processes. In section \ref{sec:full} a full cycle consisting of two isothermal and two adiabatic strokes is considered. We first investigate the quasi-static limit in which the piston speed is small and then turn to the general case of arbitrary piston speed including a discussion of efficiency at maximum power. Finally, we summarize our results in section \ref{sec:concl}.
%
\section{The model}
\label{sec:model}

The model consists of a one dimensional cylinder of length $L$ closed on top by a piston moving at constant speed $u$. The cylinder contains one classical particle with mass $m=1$ that moves with velocity $v$. At the piston the particle is assumed to be reflected elastically. With the piston being much heavier than the particle the velocity of the particle is changed from $v$ to $2u-v$ upon a reflection whereas the piston speed remains unchanged. The work done on the system by one collision of the particle at the piston is given by the change of the particles kinetic energy
\begin{equation}
  \label{eq:W}
  \Delta W = -2u(v-u).
\end{equation}

In case of an {\em adiabatic} process \cite{Lua2005} the particle is reflected elastically at the bottom of the cylinder as well. Consequently there is no heat exchange in this case. To describe an {\em isothermal} compression or expansion we thermalize the particle in the collision with the bottom of the cylinder, i.e. upon reflection we draw a new random  velocity from the distribution \cite{Baule2006,IzOk1}
\begin{equation}
  \label{eq:Phi}
  \phi(v) = \beta \,v\, e^{-\frac{\beta v^2}{2}} \,\mathds 1 [v>0]\; .
\end{equation}
Here Boltzmann's constant is set to unity, $\beta$ denotes the inverse temperature and the indicator function $\mathds 1 [\cdot]$ is one if the argument is true and zero otherwise. At the bottom the particle therefore exchanges heat with the attached reservoir but no work is performed. The distribution \eqref{eq:Phi} is chosen such that, for a static piston, $u=0$, the equilibrium Maxwell distribution is preserved. For constant $\beta$ we call the process  {\em ``isothermal''}, although, for a moving piston, in particular with large $u$, it is in general impossible to associate a temperature with the gas particle. 

%
\section{Isothermal expansion and compression}
\label{sec:isoth}

\subsection{General analysis}
\label{sec:arb_speed}

In this section, we discuss an isothermal change from the initial cylinder length $L_0$ to the final length $L_\text{f}=L_0+u T$ during time $T$ with constant speed $u$ of the piston. At time $t=0$ the system is considered to be in thermal equilibrium, i.e., the distribution of initial position $x_0$ and initial velocity $v_0$ of the particle is given by
\begin{equation}
  \label{eq:p_0}
  \rho_0(x_0,v_0) = \frac{1}{L_0}\sqrt{\frac{\beta}{2 \pi}} \exp \left(-\beta \frac{v_0^2 }{2}\right)
  \,\mathds 1[0 < x_0 < L_0].
\end{equation}

\begin{figure}
  \centering
  \includegraphics{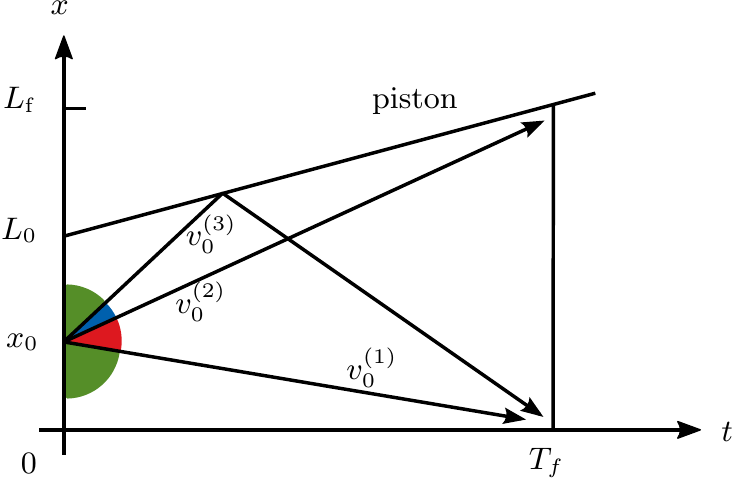}
  \caption{Classification of particle trajectories on the basis of the initial position $x_0$ and the initial speed $v_0$. For either $v_0 < v_0^{(1)}$ or $v_0 > v_0^{(3)}$ the particle reaches the bottom, $x=0$, of the cylinder (green sectors). For $v_0^{(1)} < v_0 < v_0^{(2)}$ it neither reaches the bottom nor the piston (red sector), whereas for $v_0^{(2)} < v_0 < v_0^{(3)}$ it reaches the piston but not the bottom (blue sector).}
  \label{fig:v123}
\end{figure}

To determine the distribution of the total work $W$ transferred during an expansion or compression, we need to know the times $t_j$ at which the particle hits the bottom. With the help of \eqref{eq:Phi} we may then determine its velocity and, using \eqref{eq:W}, also the work performed at the piston. If the particle leaves the bottom at time $t_j$ with velocity  $v_j$, it reaches the bottom again after the time increment $\Delta t_j = {2L_j}/({v_j-2u})$. Here $L_j=L_0+u t_j$ is the length of the cylinder at time $t_j$.  The time interval $\Delta t_j$ not only depends on $v_j$ but, through $t_j$, also on {\em all previous} velocities $v_i, i=0,..,(j-1)$ which makes further analytical progress hard \cite{Baule2006}. To disentangle this recursive dependence we introduce the {\em logarithmic} time variable
\begin{equation}
  \label{eq:tau_definition}
  \tau(t) = \ln\left(1+t\frac{u}{L_0} \right)\; .
\end{equation}
For an expanding piston, the logarithmic time increases with real time, while it decreases for a compression. The logarithmic time increments between collisions of the particle at the bottom are then given by
\begin{equation}
  \label{eq:delta_tau}
   \Delta \tau_j  =
   \begin{cases}
     \infty & \text{ for } v_j \leq 2u,\\
      \tau(t_{j+1}) -\tau(t_j) = \ln \dfrac{v_j}{v_j-2u}& \text{ otherwise}.
   \end{cases}
\end{equation}
For $v_j\leq 2u$, the velocity of the particle after the collision with the piston is $2u - v_j \geq 0$ such that the particle does not return to the bottom at all. We therefore set $\Delta\tau_j = \infty$ for this case. Since the velocities $v_j$ after successive collisions with the bottom are independent identically distributed random variables, this holds true for the time increments $\Delta\tau_j$ as well. This fact will allow us to use methods from the theory of continuous time random walks \cite{MontrollWeiss65,Balescu97} to determine the desired work distribution.  

In order to find the pdf $p(W;\tau)$ that during a process of duration $\tau=\ln L_f/L_0$ the work $W$ is performed we have to distinguish three types of particle trajectories (cf. Fig.\ \ref{fig:v123} for the case of an expansion): 
\begin{enumerate}
\item The ``null'' case in which the particle neither hits the piston nor the bottom. This happens with probability $P_0(\tau)$ that the initial velocity is between $v_0^{(1)}= {-x_0}/{T}$ and $v_0^{(2)} = {(L_0-x_0)}/{T}+u$. Since no work is performed in this case we have for the corresponding pdf $p_\text{n}(W;\tau)=\delta(W) P_0(\tau)$. 
\item The ``piston'' case in which the particle reaches the piston and performs work but does not hit the bottom. It occurs if the initial velocity lies between $v_0^{(2)}$ and $ v_0^{(3)}= {(2L_0-x_0)}/{T}+2u$. The corresponding pdf will be denoted by $p_\text{p}(W;\tau)$.
\item The ``bottom'' case in which the particle hits the bottom at least once and therefore  
gets thermalized. This happens if the initial velocity $v_0$ is either smaller than $v_0^{(1)}$ or larger than $ v_0^{(3)}$. We denote the pdf for this case by $p_\text{b}(W;\tau)$. 
\end{enumerate}

The total probability density that work $W$ is performed in the time interval $\tau$ is therefore given by

\begin{equation}
  \label{eq:total_prob}
  p(W;\tau) = \delta(W)P_0(\tau) + p_\text{p}(W;\tau) + p_\text{b}(W;\tau) \; .
\end{equation}

The simplest case to analyze is the first one. From \eqref{eq:p_0} we find 
\begin{multline}
  \label{eq:p0}
  P_0(\tau)
  = \int_0^{L_0} \text dx_0 \int_{v_0^{(1)}}^{v_0^{(2)}} \text dv_0\, \rho_0(x_0,v_0) \\
  = \frac{e^\tau-1}{u\sqrt{2 \beta}} 
  \left[
    A \erf (A) + \frac{1}{\sqrt \pi} e^{-A^2}  -B \erf (B)
  \right. \\ \left.
    - \frac{1}{\sqrt \pi} e^{-B^2}
    +C \erf (C) + \frac{1}{\sqrt \pi} e^{-C^2} - \frac{1}{\sqrt \pi}
  \right]
\end{multline}
with the constants 
\begin{equation}
 A =\sqrt{\frac{\beta}{2}}u,\quad 
 B = \sqrt{\frac{\beta}{2}} \frac{L_\text{f}\,u}{L_0(e^\tau-1)},\quad 
 C =\sqrt{\frac{\beta}{2}} \frac{u}{e^\tau-1}\; .
\end{equation}

In the second case the work $W=-2u(v_0-u)$ is performed. We hence get 
\begin{multline}
  \begin{split}
    \label{eq:p_piston} 
    p_\text{p}^{}(W;\tau) =& \int_{0}^{L_0} \!\!\text dx_0
    \int_{v_0^{(2)}}^{v_0^{(3)}}\!\!\text dv_0\
    \rho_0(x_0,v_0)\,\delta[W + 2u (v_0-u)]\\
    =& \frac{1}{|2u|} \sqrt{ \frac{\beta}{2\pi} } 
    \exp\left[
      -\beta\left(u-\frac{W}{2u} \right)
    \right]\\
    &\times \left\{
      \min \left[(e^\tau-1) \left(1-\frac{W}{2u^2} \right)
        -2,1 
      \right] 
    \right.
  \end{split} \\ \left.
      -\max\left[
        1 + (e^\tau-1)\frac{W}{2u^2},0 
      \right] 
    \right\}.
\end{multline}
The most complex case is the last one. It consists itself of three ingredients: the initial period up to the first collision with the bottom, the middle period of repeated collisions with the bottom, and the final period following the last collision with the bottom. 

Let us start with the determination of the joint pdf $p_\text{i}(\Delta W_0, \tau_0)$ for the duration and work of the first period. The logarithmic time $\tau_0$ of the first contact of the particle with the bottom is given by
\begin{equation}
  \label{eq:tau_1}
  \tau_0 = 
  \begin{cases}
    \ln\left(1-\frac{x_0 u}{L_0 v_0} \right) 
      &\text{ for } v_0 < \min\left(0, \frac{x_0 u}{L_0}\right),\\
    \ln\left(1 +\frac{(2L_0 - x_0)u}{ (v_0-2u)L_0} \right)
      &\text{ for } v_0 > \max \left(  2u, \frac{x_0 u}{L_0}  \right).
  \end{cases}
\end{equation}

The values of $v_0$ not covered above are those realizations considered  by  $P_0(\tau)$ and $p_\text{p}(W;\tau)$.
In the first case of \eqref{eq:tau_1} the particle  moves directly to the bottom and hence does not transfer work before the first contact with the bottom.
In the second case it reaches the piston first  and does transfer work before hitting the bottom.  From \eqref{eq:p_0} we  get

\begin{widetext} 
  \begin{equation}
    \begin{split}
    \label{eq:p_initial}
    p_\text{i}(\Delta W_0,\tau_0)
    =& \delta(\Delta W_0) \int_0^{L_0} \mathrm dx_0 \int_{-\infty}^\infty \mathrm dv_0\, \rho(x_0,v_0) 
    \,\delta\left[\tau_0 -  \ln\left(1-\frac{x_0 u}{L_0 v_0} \right) \right]\,
    \mathds 1 \left[
      v_0 < \min\left(0, \frac{x_0 u}{L_0}\right)
    \right] \\
    &+  \int_0^{L_0} \mathrm dx_0 \int_{-\infty}^\infty \mathrm dv_0\, \rho(x_0,v_0)
    \delta[W_0 +2u(v_0-u)]  \,
    \delta\left[
      \tau_0 - \ln\left(1 +\frac{(2L_0 - x_0)u}{ (v_0-2u)L_0} \right)
    \right]
      \mathds 1 \left[
      v_0 > \max\left(2u, \frac{x_0 u}{L_0}\right)
    \right] \\
    =& \frac{e^{\tau_0}}{|u| \sqrt{2\pi}}
    \left\{ \frac{\delta(\Delta W_0)}{\sqrt{\beta}} 
      \left[ 1- 
        \exp \left(
          -\frac{\beta}{2} \left(\frac{u}{e^{\tau_0}-1}\right)^2
        \right)
      \right] 
    \right.\\
    &\left.
      +\frac{\sqrt\beta}{2} 
      \,\left|  \frac{\Delta W_0}{2u^2} +1 \right|
      \exp\left[
        -\frac{\beta}{2}\left( u - \frac{\Delta W_0}{2u} \right)^2
      \right]
      \mathds 1[0<\chi(\Delta W_0,\tau_0)<1]  
    \right\}
    \mathds 1 [\tau_0>0]
  \end{split}
  \end{equation}
\end{widetext}
with  $\chi(\Delta W_0,\tau_0) = 2 + (e^{\tau_0}-1)[\Delta W_0/(2u^2)+1]$.

Note that $\int \text d \Delta W_0\, \text d\tau_0\, p_\text{i}(\Delta W_0,\tau_0) \le 1$ since the particle never reaches the bottom if its initial velocity is between zero and $2u$. This case is covered by $P_0(\tau)$ and $p_\text{p}(W;\tau)$ respectively.
 
To characterize the middle period of repeated collisions we introduce the pdf 
$\psi(\Delta W, \Delta\tau)$ giving the joint probability density that between two collisions of the particle with the bottom the logarithmic time increases by $\Delta\tau$ and the work $\Delta W$ is performed. It is most conveniently specified by its Fourier transform in $W$ and (bilateral) Laplace transform in $\Delta \tau$
\begin{equation}
 \hat{\tilde \psi}(k,\lambda) 
  =\int_{-\infty}^\infty \!\!\text d \Delta \tau \int_{-\infty}^\infty \!\!\text d \Delta W\, e^{-\lambda \Delta \tau +i k\Delta W}
      \psi(\Delta W,\Delta\tau).
\end{equation}
Since $\tau$ is either positive for expansions and negative for compression we have to use the bilateral Laplace transform in contrast to the usual formalism which uses an unilateral Laplace transform \cite{Balescu97}. The Laplace transform is defined for $\Re \lambda >0$ for $u>0$ and $\Re \lambda < 0$ for $u < 0$. Using \eqref{eq:W}, \eqref{eq:delta_tau}, and \eqref{eq:Phi} it acquires the explicit form 
\begin{multline}
  \label{eq:phi_laplace-fourier}
\begin{split}
  \hat{\tilde \psi}(k,\lambda) 
  &= \int_{0}^\infty \text dv\, \phi(v) 
  e^{
    -\lambda \Delta\tau(v) + i k\Delta W(v)
  } \\
  &= \int_{\max(2u,0)}^\infty \text dv\, \phi(v) \left( 1 - 2 \frac{u}{v} \right)^\lambda e^{-ik2u(v-u)} \; .
\end{split}
\end{multline}
The integration starts from $\max(2u,0)$ since $\Delta \tau(v) = \infty$ for $v<2u$ (Eq.~\eqref{eq:delta_tau}). For $u > 0$ we have $\lim_{\lambda \to \infty} \hat{\tilde {\psi}}(0,\lambda) < 1$ which is the probability that the particle returns to the bottom.

The period after the last collision with the bottom is characterized by the fact that the time $\Delta\tau_\text{f}$ remaining till the end of the process is not sufficient for the particle to reach the bottom again. To determine the final contribution to the work we have to distinguish whether the particle reaches the piston a last time or not. If  $\Delta \tau_\text{f}<\ln[v/(v-u)]$ it does not and there is no final work increment (first term in \eqref{eq:pf}). If $\ln[v/(v-u)]<\Delta \tau_\text{f}<\ln[v/(v-2u)]$ the particle hits the pistons again and performs a final contribution $\Delta W_\text{f}$ to the work (second term in \eqref{eq:pf}). Note that $\Delta \tau_\text{f}$ cannot exceed $\ln[v/(v-2u)]$ according to \eqref{eq:delta_tau}.

\begin{widetext}
  \begin{equation}
    \label{eq:pf}
    \begin{split}
    p_\text{f}(\Delta W_\text{f};\Delta \tau_\text{f}) =& 
    \delta(\Delta W_\mathrm{f}) \int_0^\infty \mathrm dv\, 
    \phi(v) \, \mathds 1\left[v < \frac{u}{1-e^{-\Delta \tau_f}} \right]\\
    &+  \int_0^\infty \mathrm dv\,  \phi(v) \delta[\Delta W_\mathrm{f} +2u(v-u)] \,
    \mathds 1 \left[
      \frac{u}{1-e^{-\Delta \tau_f}} < v <  \frac{2u}{1-e^{-\Delta \tau_f}}
    \right]\\
    =& \delta(\Delta W_\text{f})
    \left\{
      1-
      \exp \left[
        -\frac{\beta}{1} \left(\frac{u }{1-e^{-\Delta \tau_\text{f}} } \right)^2
      \right]
    \right\}
    +
    \frac{1}{|2u|}
    \phi \left(
      u - \frac{\Delta W_\text{f}}{2u} 
    \right)
    \mathds 1 \left[
      \frac{u}{1-e^{-\Delta \tau_f}} < u - \frac{\Delta W_\mathrm{f}}{2u} <  \frac{2u}{1-e^{-\Delta \tau_f}}
    \right].
  \end{split}
  \end{equation}
\end{widetext}
Now, assume that we start for $\tau=0$ at the bottom and denote the pdf for this process by $p_{\mathrm{s}}(W;\tau)$. When the particle hits the bottom again, the process can be described by the same distribution $p_{\mathrm{s}}(W;\tau)$ from this point on (also called a \emph{renewal}) \cite{MontrollWeiss65}. This leads to the equation
\begin{multline}
\label{eq:consistency}
p_{\mathrm{s}}(W;\tau) 
= p_{\mathrm{f}}(W;\tau)\\
 + \int_{-\infty}^\infty \mathrm{d}W' \, \int_{-\infty}^\infty \mathrm{d}\tau' \, \psi(W',\tau') p_{\mathrm{s}}(W-W'; \tau - \tau').
\end{multline}
This relation is quite intuitive: the first part describes the case that we do not hit the bottom again, the second part describes the renewal property which is a process of remaining length $\tau - \tau'$ which already has acquired a work contribution of $W'$ in advance. Taking the Fourier transform with respect to $W$ and the Laplace transform with respect to $\tau$ turns Eq.~\eqref{eq:consistency} into the algebraic equation
\begin{equation}
\hat{\tilde{p}}_{\mathrm{s}}(k;\lambda) = \hat{\tilde{p}}_{\mathrm{f}}(k;\lambda) + 
  \hat{\tilde{\psi}}(k,\lambda) \hat{\tilde{p}}_{\mathrm{s}}(k;\lambda)
\end{equation}
which gives
\begin{equation}
\label{eq:FL-simplified-process}
\hat{\tilde{p}}_{\mathrm{s}}(k;\lambda)
= \frac{\hat{\tilde{p}}_{\mathrm{f}}(k;\lambda)}{1 - \hat{\tilde{\psi}}(k,\lambda)}.
\end{equation}
We are now in a position to derive an equation for the overall $p_\text{b}(W;\tau)$ characterizing the bottom case. We have to accommodate the fact that we do not necessarily start from the bottom, but the first hit is described by the pdf $p_{\mathrm{i}}(W,\tau)$. Along the lines of the renewal argument, we get the convolution
\begin{equation}
p_\text{b}(W;\tau)
= \int_{-\infty}^\infty \text dW'\, \int_{-\infty}^\infty \text d\tau'\, p_\text{i}(W',\tau')\,p_\text{s}(W-W';\tau-\tau').
\end{equation}

Taking again the Fourier transform with respect to $W$ and the Laplace transform with respect to $\tau$ gives
\begin{equation}
  \label{eq:p_laplace_fourier}
  \hat{\tilde p}_\text{b}(k;\lambda) = 
  \frac{
    \hat{\tilde p}_\text{i}(k,\lambda)\,
    \hat{\tilde p}_\text{f}(k;\lambda)
  }{
    1-\hat{\tilde \psi}(k,\lambda)
  }.
\end{equation}
Upon inverse Fourier and Laplace transformation we get the result for $p_\text{b}(W;\tau)$ which completes the analysis of the bottom case and fixes the last term in \eqref{eq:total_prob}. 
Unfortunately neither the $v$-integral in \eqref{eq:phi_laplace-fourier} nor the inverse transformation of \eqref{eq:p_laplace_fourier} can be performed analytically. However, they can be done efficiently using numerical methods.

\begin{figure*}
  \centering
  \includegraphics{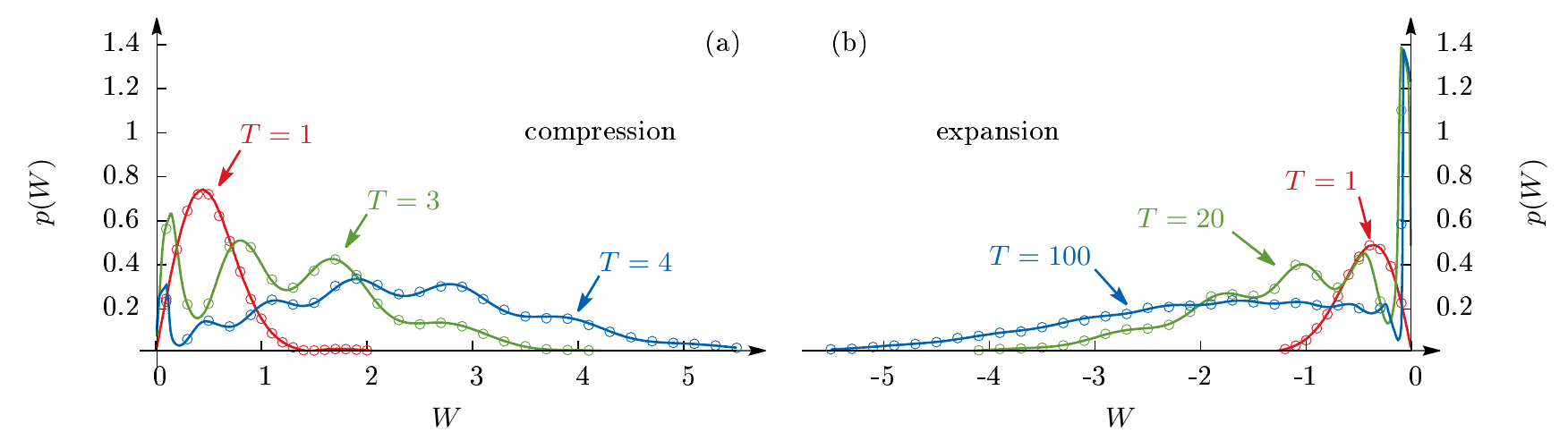}
  \caption{Time evolution of the distribution of work $p(W;\tau)$ for different durations $T=L_0(e^\tau-1)/u$, for both compressions (a)  and expansions (b). The system starts in equilibrium at temperature $1/\beta=1$, the initial length of the cylinder is set $L_0=1$, the piston speed is $u=\pm 0.2$. Shown are the analytic expressions [lines, eq.\ \eqref{eq:total_prob}] together with results of molecular dynamics simulations (circles) of $10^6$ trajectories each. The $\delta$-peaks at $W=0$ are not displayed.}
  \label{fig:time_evolution}
\end{figure*}

\begin{figure*}
  \centering
  \includegraphics{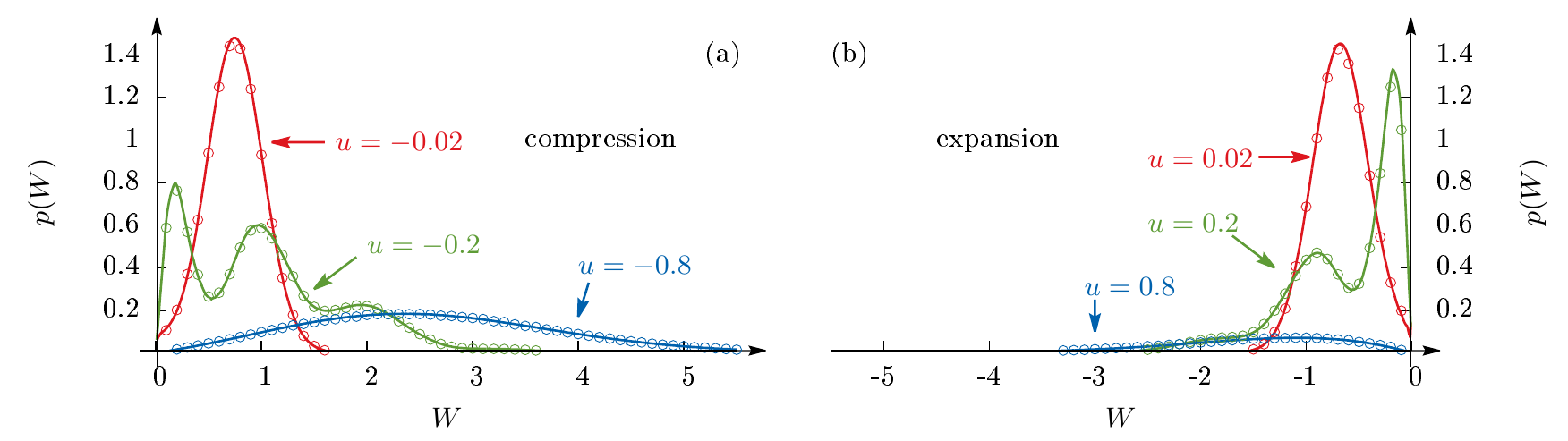}
  \caption{Distribution of work $p(W;\tau)$ for compressions (a) and expansions (b) starting from equilibrium at temperature $1/\beta=1$ for different speeds $u$ of the piston. The length of the cylinder is changed from $L_0=1$ to $L_\text{f}= 2$ and vice versa respectively. Shown are the analytic expressions [lines, eq.\ \eqref{eq:total_prob}] together with results of molecular dynamics simulations (circles) of $10^6$ trajectories each. The $\delta$-peaks at $W=0$ are not displayed.}
  \label{fig:u_dependence}
\end{figure*}

In Figs.\ \ref{fig:time_evolution} and \ref{fig:u_dependence} we show several examples of work distributions obtained from \eqref{eq:total_prob} [inserting Eqs.~\eqref{eq:p0}, \eqref{eq:p_piston}, \eqref{eq:p_laplace_fourier}, \eqref{eq:p_initial},  \eqref{eq:phi_laplace-fourier} and  \eqref{eq:pf}] and compare them with results from numerical simulations of the particle dynamics. The agreement between analytical and numerical results is always very good. The relative deviations are typically below 1\%. 

Fig.\  \ref{fig:time_evolution} shows the evolution of $p(W;\tau)$ with time. For both compression and expansion the probability that the particle has hit the piston more than once is small for small times $T$. The distribution of work therefore shows only one major peak. With increasing $T$, also the probability for multiple collisions increases, resulting in a  distribution with several peaks. At very long times, these peaks begin to broaden and finally merge. 

The distributions for expansion and compression mainly differ in the time evolution of the peaks at small work values. In an expansion the probability for the particle to hit the piston only once does not decrease with time and hence the height of the peak in  $p(W,\tau)$ at small $W$ remains roughly constant, cf. the curves for $T=20$ and $T=100$ in Fig.\  \ref{fig:time_evolution} (b). For a compression the steadily decreasing volume reduces the probability for just a single collision between particle and piston and correspondingly the peak at small values of $W$ gradually disappears, as shown in Fig.\  \ref{fig:time_evolution} (a).

\begin{figure}
  \centering
  \includegraphics{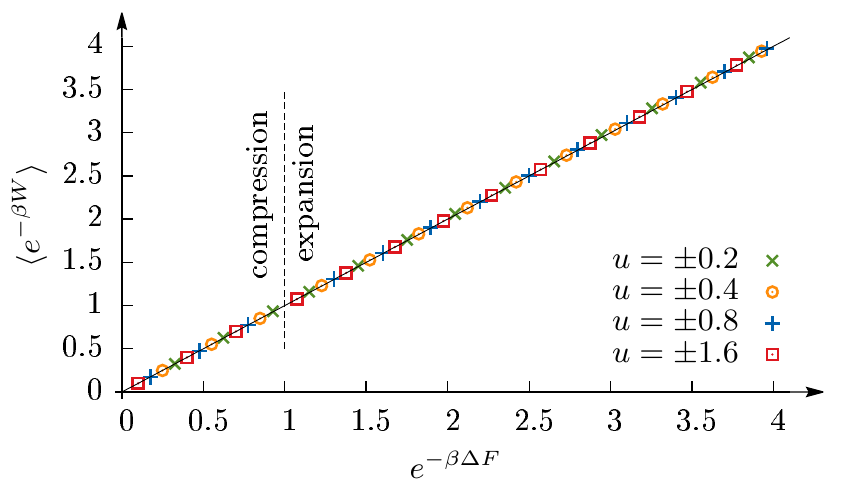}
  \caption{Verification of the Jarzynski equality. The mean $\langle e^{-\beta W}\rangle$  for compressions and expansions starting from $L_0 =1$ in equilibrium at temperature 1 is determined for different speeds of the piston and different final lengths of the cylinder. The results are plotted against $e^{-\beta \Delta F}$ which is proportional to the final length $L_\text{f}$.}
  \label{fig:jarzynski}
\end{figure}

The influence of the piston speed $u$ on the work distributions is shown in Fig.\ \ref{fig:u_dependence}. For a very slow piston there are many collisions with the particle, the work increments of the individual collisions are independent random variables and by the central limit theorem $p(W,\tau )$ becomes Gaussian. For intermediate piston speeds, the distribution develops multiple peaks originating from realizations with different
numbers of collisions. For high piston speed, $p(W,\tau)$ is dominated by a peak at $W=0$ (not shown in the Figs.) stemming from realizations where the particle does not reach the piston at all and a broad unimodal contribution originating from just a single collision between particle and piston.

The main difference between the distributions shown in Figs.\ \ref{fig:time_evolution} and \ref{fig:u_dependence} and those obtained for adiabatic compression and expansion respectively \cite{Lua2005} lies in the long-time behavior. Whereas in the adiabatic case the features resulting from individual collisions leave their trace in $p(W;\tau)$ for arbitrary long times they are washed out in the isothermal case as time goes by. This is consistent with the fact that the only random variables in the former case are those from the initial condition whereas in the isothermal process a new random variable enters the stage after each collision with the bottom. 

To check our results for $p(W;\tau )$ we have tested them against the Jarzynski identity \cite{Jar97} and the Crooks relation \cite{Crooks}. The Jarzynski equation stipulates  $\langle e^{-\beta W}\rangle=e^{-\beta\Delta F}$ with the free-energy change
\begin{equation}\label{eq:defF}
 \Delta F=-\frac{1}{\beta}\ln\frac{L_\text{f}}{L_0}=-\frac{\tau}{\beta}\; .
\end{equation}   
Fig.\  \ref{fig:jarzynski} demonstrates that the average $\langle e^{-\beta W}\rangle$ determined with various distributions $p(W;\tau )$ obtained from \eqref{eq:total_prob} indeed coincides with $e^{-\beta\Delta F}$ of the corresponding process. The deviations are less than 1\%  and can be assigned to round-off errors in the numerics for the negative tails of $p(W;\tau)$.

\begin{figure}
  \includegraphics{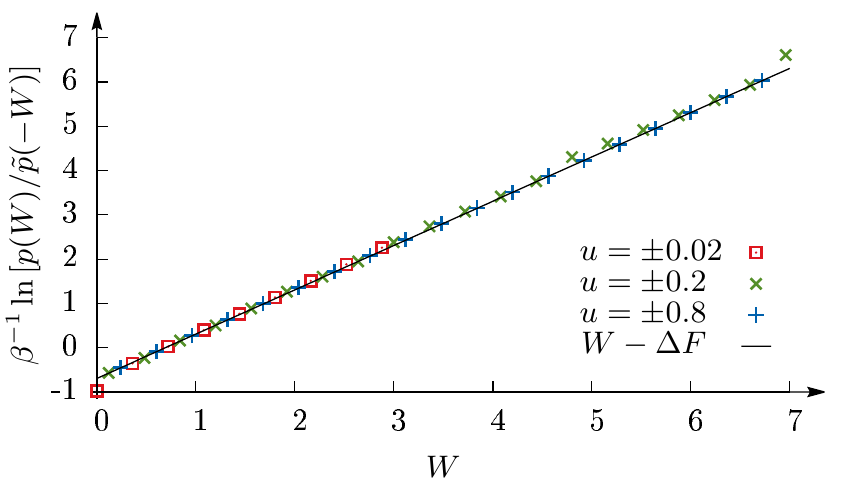}
  \caption{Verification of the Crooks relation. The log-ratio of the work distribution $p(W)$ of an isothermal compression from $L_0 = 2$ to $L_\text{f} = 1$  and  $\tilde p(-W)$ of the corresponding reverse process is plotted for different values of $u$. All points lie on the line $W-\Delta F$ as required by \eqref{eq:crooks}. For $|u| = 0.02$ and $W > 3$, $p(W)$ and $\tilde p(-W)$ are less than $10^{-16}$ and dominated by round-off errors. Hence these values are not shown.}
  \label{fig:crooks}
\end{figure}

The Crooks relation 
\begin{equation}
  \label{eq:crooks}
  \frac{p(W;\tau)}{ \tilde p(-W;\tau)} = e^{\beta (W-\Delta F)}.
\end{equation}
involves the distribution of work for the reverse process, $\tilde{p}(W;\tau)$. The reverse process of an expansion (compression) from $L_0$ to $L_\text{f}$ at piston speed $u$ is a compression (expansion) form $L_\text{f}$ to  $L_0$ at piston speed $-u$. 
In Fig.\ \ref{fig:crooks} the logarithmic ratio $ \beta^{-1} \ln (p(W;\tau)/\tilde p(-W;\tau))$ is plotted against $W$. Deviations from \eqref{eq:crooks} are only visible where $p(W;\tau)$ or $\tilde p(-W;\tau)$ are very small (less than $10^{-4}$) and can again be attributed to round-off errors.

\subsection{Quasi-static limit}
\label{sec:quasi-static-one_step}

While it seems not possible to get an explicit analytic expression for Eq.~\eqref{eq:total_prob}, it is feasible to get results in the limit $u \to 0$. In order to shorten the derivation we make the further assumption that the particle starts at the
bottom of the cylinder. The error resulting for the work is of order $u$ and can be neglected.

For this process the probability $p_{\mathrm{s}}(W;\tau)$
can be expressed in Fourier-Laplace space as (Eq.~\eqref{eq:FL-simplified-process})
\begin{equation}
\label{eq:simplified_probability}
\hat{\tilde{p}}_\mathrm{s}(k;\lambda) = \frac{\hat{\tilde{p}}_\mathrm{f}(k;\lambda)}{1-\hat{\tilde{\psi}}(k,\lambda)}.
\end{equation}
With the definition
\begin{equation}
\label{eq:def_gu}
\begin{split}
g_u(k,\lambda) 
&= \int_{\max(u,0)}^\infty \text dv\, \phi(v) 
    \left(1 - \frac{u}{v}\right)^\lambda \exp(- i k u v) \\
&= 1 - u \sqrt{\frac{\pi\beta}{2}} \lambda - i u \sqrt{\frac{\pi}{2 \beta}} k + o (u)
\end{split}
\end{equation}
we have from Eq.~\eqref{eq:phi_laplace-fourier}
\begin{equation}
\begin{split}
\hat{\tilde{\psi}}(k,\lambda)
&= e^{2 i k u^2} g_{2u}(k,\lambda) \\
&= 1 - u \sqrt{2\pi\beta} \lambda - i u \sqrt{\frac{2 \pi}{\beta}} k + o (u)
\end{split}
\end{equation}
and from Eq.~\eqref{eq:pf}
\begin{equation}
\begin{split}
\hat{\tilde{p}}_\mathrm{f}(k,\lambda)
&= \frac{1}{\lambda} \left( 1 - g_u(0,\lambda) + e^{2iku^2} g_u(2k,\lambda) - \hat{\tilde{\psi}}(k,\lambda) \right) \\
&= u \sqrt{2\pi\beta} + o(u).
\end{split}
\end{equation}
Plugging this into Eq.~\eqref{eq:simplified_probability} results in

\begin{equation}
\hat{\tilde{p}}_\mathrm{s}(k,\lambda)
= \frac{1}{\lambda + i k \beta^{-1}}
+ o(1).
\end{equation}

The inverse Fourier-Laplace-transform gives the final result
\begin{equation}
  \label{eq:W_quasi-static}
\begin{split}
  p_\text{s}(W;\tau) 
    &= \delta(W + \beta^{-1} \tau) \\
    &= \delta(W-\Delta F).
\end{split}
\end{equation}
This result is calculated for the simplified process $p_\text{s}(W;\tau)$, however the corrections to $p(W;\tau)$ is only the behavior till the first hit at the bottom which is of lower order than considered here. Hence, in the quasi-static limit and to leading order, the work does not fluctuate and is always equal to the free-energy change $\Delta F$. This result is also suggested by the law of large numbers since the work is performed in infinitely many collision between the particle and the piston. In contrast, during an adiabatic compression or expansion \citep{Lua2005} the individual collisions stay correlated and the work fluctuates even in the quasi-static limit.
%
\section{The full cycle} 
\label{sec:full}
%

\begin{figure}
  \includegraphics{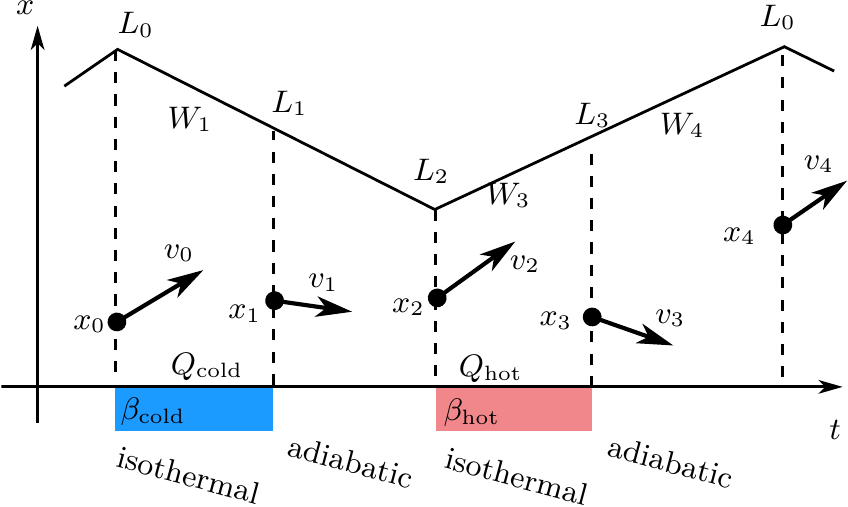}
  \caption{Variables used to describe the full cyclic process discussed in section \ref{sec:full}. $x_i$ and $v_i$ denote the position and the speed of the particle at the end of the $i$-th stroke, $L_i$ is the length of the cylinder at this moment, and $W_i$ and $Q_i$ denote the work and heat respectively exchanged during the $i$-th stroke.}
  \label{fig:cycle_variables}
\end{figure}

We now combine our results of section \ref{sec:arb_speed} for isothermal volume changes with those of \citet{Lua2005} for adiabatic ones to analyze the fluctuations of work and heat in a full cyclic process consisting of four strokes, similar to a Carnot process. The organization of the cycle together with the relevant notations is shown in Fig.~\ref{fig:cycle_variables}. We choose the cycle to start with an isothermal compression at inverse temperature $\beta_\text{cold}$. It is followed by an adiabatic compression changing the inverse temperature from  $\beta_\text{cold}$ to $\beta_\text{hot}$. The third stroke is an isothermal expansion at the higher temperature followed by an adiabatic expansion back to  initial volume. This definition of the cycle is somewhat arbitrary; the cycle could start at any stroke. As before the initial position and velocity of the particle are denoted by $x_0$ and $v_0$ respectively, $x_i$ and $v_i$ are their respective values after the $i$-th stroke. The length of the cylinder at these times is denoted by $L_i$. To keep the analysis simple we choose the same piston speed $u_\text{c}$ for both compression strokes and the same value $u_\text{e}$ for both expansions. 

The heat transferred during the isothermal strokes can be determined by applying the first law of thermodynamics to each individual realization of the process. Since the change in internal energy is just the change in kinetic energy of the particle, we find
\begin{align} 
  \label{eq:heat_quasi_stat_cycle1}
  Q_\text{cold} &= \frac{1}{2}\left(v_1^2 -v_0^2\right) -W_1,\\ \label{eq:heat_quasi_stat_cycle2}
  Q_\text{hot} &= \frac{1}{2}\left(v_3^2 -v_2^2 \right) -W_3.
\end{align}

A general analysis of the full cycle is hampered by the fact that the distributions of $x_i$ and $v_i$ at the beginnings of the different strokes are not known. Note that it is not even acceptable to use \eqref{eq:Phi} for the initial distribution of $(x_0,v_0)$, since in general we would not come back to it after the fourth stroke, and would therefore not describe a time-periodic steady state (TPSS). We hence start with an investigation of the quasi-static limit of the cycle in which the situation is somewhat simpler and some comparison between analytical work and simulations is still possible. After this we elucidate the general case with arbitrary values of $u_\text{c}$ and $u_\text{e}$ on the basis of numerical simulations. 

\subsection{Quasi-static case}
\label{sec:quasi-static_cycle}

\begin{figure*}
  \includegraphics{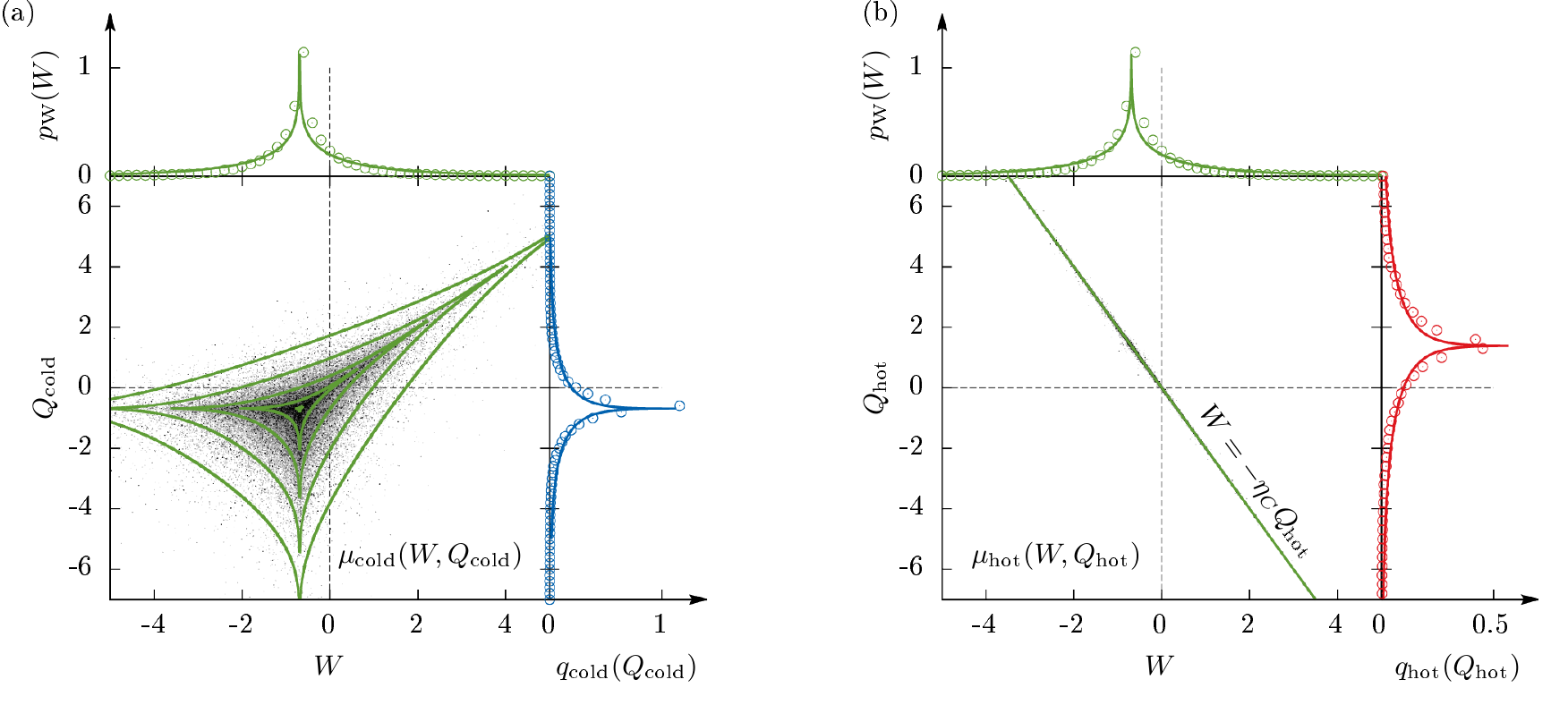}
  \caption{Comparison of the analytical solutions for the quasi-static case with simulations of  $10^5$ cycles with a piston speed $u=\pm10^{-5}$.  In the subplots, the respective joint pdfs $\mu_\text{cold}(W,Q_\text{cold})$ (a) and $\mu_\text{hot}(W,Q_\text{hot})$ (b) of the work $W$ and the heat $Q_i$ are shown in the center. The analytical solution is depicted by lines of equal probability, while each simulated cycle is represented by one black dot. The marginal probability $p_W(W)$ is shown on the top of each subplot whereas the marginal probabilities $q_i(Q_i)$ can be seen at the right hand sides of (a) and (b). Here lines denote the analytical solutions, while the simulations are depicted as circles. The inverse temperatures are $\beta_\text{hot}^{-1}= 2$ and $\beta_\text{cold}^{-1}= 1$, and in the isothermal stroke at the higher temperature the cylinder length changes from $L_2=1$ to $L_3=2$.}  
  \label{fig:quasi_stat}
\end{figure*}

The quasi-static case is defined by the combined limit $u_\text{c} \rightarrow 0$ and $u_\text{e} \rightarrow 0$. From section \ref{sec:quasi-static-one_step} we know that in these limits the work performed in the isothermal strokes equals the free-energy difference, cf. \eqref{eq:W_quasi-static}, i.e.
\begin{align}\label{eq:W_1}
 W_1 = \Delta F_\text{cold} &= -\frac{1}{\beta_\text{cold}}\ln \frac{L_1}{L_0}\\ \label{eq:W_3}
 W_3 = \Delta F_\text{hot}  &= -\frac{1}{\beta_\text{hot}}\ln \frac{L_3}{L_2}\; .
\end{align} 

In the adiabatic strokes the number of collisions $n_i$ is given for small piston speed $u$ by \cite{Lua2005}
\begin{equation}
  \label{eq:n_adiabatic}
  n_i = \frac{|v_i|}{2u}\left(1-\frac{L_i}{L_{i+1}} \right) + O(1).
\end{equation}
 After these collisions the particle speed is reduced to 
\begin{equation}
\label{eq:v_f_adiabatic}
 |v_{i+1}| = |v_i|-2n_i u  = |v_i|\frac{L_i}{L_{i+1}} + O(u)
\end{equation}
and the work performed is therefore given by 
\begin{multline} 
  \label{eq:W_adiabatic}
  W_{i+1} =  - 2n_i u(|v_i|-n_i u) 
  =  \frac{v_i^2}{2} \left(\frac{L_i^2}{L_{i+1}^2} -1 \right) + O(u)\\
\end{multline}

From \eqref{eq:v_f_adiabatic} it follows that if $v_i$ obeys a Maxwell distribution at inverse temperature $\beta_i$ then $v_{i+1}$ will be Maxwell distributed as well, however at inverse temperature 
\begin{equation}
  \label{eq:beta_f}
  \beta_{i+1} = \beta_i\frac{L_{i+1}^2}{L_i^2}\; ,
\end{equation}
as expected for a classical ideal gas in one dimension. Since \eqref{eq:Phi} ensures that the Maxwell distribution is reproduced after each isothermal collision we find back the well-known fact that in the quasi-static limit the system is always in equilibrium. 

Hence
\begin{align}
  \label{eq:L_restictions}
 \frac{L_1}{L_2}=\frac{L_0}{L_3}=\sqrt{\frac{\beta_\text{cold}}{\beta_\text{hot}}}
\end{align}
and we have to leading order
\begin{align}\label{eq:W_2}
 W_2 &= \frac{v_1^2}{2}\left(\frac{\beta_\text{cold}}{\beta_\text{hot}}  -1\right)\\
 \label{eq:W_4}
 W_4 &= \frac{v_3^2}{2}\left(\frac{\beta_\text{hot}}{\beta_\text{cold}} -1\right)\; .
\end{align} 

From \eqref{eq:heat_quasi_stat_cycle1}, \eqref{eq:heat_quasi_stat_cycle2}, \eqref{eq:W_1}, \eqref{eq:W_3}, \eqref{eq:W_2}, and \eqref{eq:W_4} it is clear that the total work $W = \sum_{i=1}^4 W_i$ as well as $Q_\text{cold}$ and $Q_\text{hot}$ depend on the three velocities $v_0$, $v_1$, and $v_3$ only. These velocities are independent samples from the Maxwell distribution. The velocity $v_2$ is then fixed by $v_2 = v_1 \sqrt{\beta_\text{cold} / \beta_\text{hot}}$.

The pdf $p(W,Q_\text{cold},Q_\text{hot})$ characterizing the whole cycle is given by
\begin{widetext}
  \begin{multline}
    p(W,Q_\text{cold},Q_\text{hot}) = \frac{\beta_\text{cold}
      \sqrt{\beta_\text{hot}}}{(2 \pi)^{3/2}}
    \int_{-\infty}^\infty \text dv_0\, e^{-\beta_\text{cold} \frac{v^2_0}{2}} 
    \int_{-\infty}^\infty \text dv_1\, e^{-\beta_\text{cold}
      \frac{v^2_1}{2}}
    \int_{-\infty}^\infty \text dv_3\, e^{-\beta_\text{hot} \frac{v^2_3}{2}} \\
    \delta \left[ W - \Delta F_\text{cold} -\frac{v_1^2}{2}\left(
        \frac{\beta_\text{cold}}{\beta_\text{hot}} -1 \right) - \Delta
      F_\text{hot} - \frac{v_3^2}{2}\left(
        \frac{\beta_\text{hot}}{\beta_\text{cold}} -1 \right)
    \right]\\ \times
    \delta \left( Q_\text{cold} - \frac{v_1^2 -v_0^2}{2} + \Delta
      F_\text{cold}
    \right)
    \delta \left( Q_\text{hot} - \frac{v_3^2
        -v_1^2\frac{\beta_\text{cold}}{\beta_\text{hot}}}{2} + \Delta
      F_\text{hot} \right)
  \end{multline}
\end{widetext}

 To disentangle this expression we introduce the characteristic function 
\begin{equation}
 \tilde p(k,k',k'') 
    = \left\langle 
      \exp\left\{
        ik W +ik'Q_\text{cold} +ik''Q_\text{hot}
      \right\} 
    \right\rangle
\end{equation} 
to get 
\begin{multline}
  \begin{split}
    \label{eq:char_fkt}
    \tilde p(k&,k',k'') \\
    =& \frac{1}{ \sqrt{1-{\beta'}^{-1}ik -
        \beta_\text{cold}^{-1}ik' + \beta_\text{hot}^{-1}ik''}
    }\\
    & \times\frac{1}{\sqrt{1+\beta^{-1}_\text{cold} ik}}
    \frac{1}{\sqrt{1 + {\beta'}^{-1}ik - \beta_\text{hot}^{-1}ik''}} \\
    & \times \exp\left\{ i \ln\left( \frac{L_1}{L_2}\right) \left( \beta'^{-1}k
        -\beta_\text{cold}^{-1}k' + \beta_\text{hot}^{-1}k'' \right)
    \right\},
  \end{split}
\end{multline}
with $\beta'^{-1} = \beta_\text{cold}^{-1} - \beta_\text{hot}^{-1}$. The inverse Fourier-transformation of \eqref{eq:char_fkt} cannot be done exactly in full generality. However, we may get analytical results for some marginal distributions. 

To begin with we consider the distributions of heat transferred in the individual strokes which we denote by $q_\text{cold}(Q_\text{cold})$ and $q_\text{hot}(Q_\text{hot})$ respectively. Their characteristic functions are given by $\tilde q_\text{cold}(k) = \tilde p(0,k,0)$ and $\tilde q_\text{hot}(k) = \tilde p(0,0,k)$ respectively. The inverse Fourier-transform of these characteristic functions can be done analytically and gives
\begin{equation} 
  \label{eq:p(Q_i)}
  q_i(Q_i) = \frac{\beta_i}{\pi}\text K_0\left[\beta_i (Q_i+\Delta F_i) \right],\quad i = \text{cold, hot},
\end{equation}
where  $\text K_0$ denotes a modified Bessel function of the second kind. A similar expression was found for the {\em equilibrium} heat fluctuations of a Brownian particle in a parabolic potential \citep{Imparato2007,Chatterjee2010}. Related is also the result of an exponential tail of the distribution of injected power for a Brownian particle in contact with two heat reservoirs as derived in \citep{Visco2006}. In \citep{VanZon2004} it was shown that the distribution of heat has exponential tails even for an overdamped Brownian particle in a {\em moving} parabolic potential, i.e. in a non-equilibrium situation.

For our system also the joint pdf $\mu_\text{hot}(W,Q_\text{hot})$ for the total work and heat exchanged with the cold reservoir can be determined exactly. Its characteristic function is given by $\tilde \mu_\text{hot}(k,k') = \tilde p (k,0,k')$ which upon inverse Fourier transformation yields 
\begin{equation}
  \label{eq:p(W,Q_3)}
  \mu_\text{hot}(W,Q_\text{hot}) = q_\text{hot}(Q_\text{hot})\,\delta(W+\eta_C Q_\text{hot}).
\end{equation}
Here $\eta_C=1-\beta_\text{hot}/\beta_\text{cold}$ is the Carnot efficiency. Integrating over $Q_\text{hot}$ we also find the marginal distribution $p_W(W)$. Finally,  $\mu_\text{cold}(W,Q_\text{cold})$ is obtained by performing the inverse Fourier transform numerically. 

In Fig.~\ref{fig:quasi_stat} $\mu_\text{cold}(W,Q_\text{cold})$ and  $\mu_\text{hot}(W,Q_\text{hot})$ are compared with  results from molecular dynamics simulations. There is good agreement, with small deviations due to the fact, that the simulations were performed with finite piston speed  $u=10^{-5}$. As can be seen, also in the quasi-static limit, in a Carnot cycle with a {\em single} particle, there remain substantial fluctuations of $W$, $Q_\text{hot}$, and $Q_\text{cold}$. This seems surprising since both work and heat are transferred in infinitely many collisions of the particle with the piston and the bottom, respectively. In fact, in the isothermal strokes the work is equal to the difference in free energy and therefore does not fluctuate. In the adiabatic strokes, however, the individual collisions with the piston are all correlated with each other, the law of large numbers does not apply, and fluctuations in the work remain. It is these fluctuations from the adiabatic strokes that entail the non-trivial structure of the distribution for the total work.

The heats exchanged with the reservoirs during the isothermal strokes are equal to the difference between the kinetic energy of the particle and the work performed, cf. \eqref{eq:heat_quasi_stat_cycle1} and \eqref{eq:heat_quasi_stat_cycle2}. In the quasi-static limit the work in the isothermal strokes does not fluctuate and the heats therefore just depend on the two random velocities of the particle at the beginning and at the end of the respective stroke. The influence of these velocities on the heats is not diminished by $u\to 0$ and consequently the central limit theorem is again not applicable. Therefore, the distributions of $Q_\text{hot}$ and $Q_\text{cold}$ are non-trivial as well. If there are many particles in the cylinder their independent contributions combine and the results of macroscopic thermodynamics are, of course, restored. 

Also the correlations between heat and work are interesting. Since $W$ and $Q_\text{hot}$ depend on the same random variables, namely $v_1$ and $v_2$, they are proportional to each other as expressed by \eqref{eq:p(W,Q_3)}.  Note that this relation also implies that the fluctuation theorem for heat engines as derived in \citep{Sinitsyn2011} is trivially fulfilled.. 

In contrast, $Q_\text{cold}$ depends in addition on $v_0$ and the correlation with $W$ is not as strong as for $Q_\text{hot}$. We need to emphasize, however, that there is some arbitrariness to these statements since they depend on where we define the beginning of our cycle. So far we have chosen the isothermal compression at the lower temperature as the first stroke, cf. Fig.~\ref{fig:cycle_variables}. If, instead, we were starting with the isothermal expansion $W$ and $Q_\text{cold}$ were proportional to each other and $W$ and $Q_\text{hot}$ would show only weak correlations. This is also the general picture for the quasi-static case: irrespective of the starting point of the cycle there is always one heat that is proportional to the work and one that shows relative fluctuations. This dependence on the precise definition of the cycle is restricted to the correlations, however, the marginal distributions of work and heat remain unchanged.

\subsection{Arbitrary piston speed and efficiency at maximum power}

\begin{figure*}
  \includegraphics{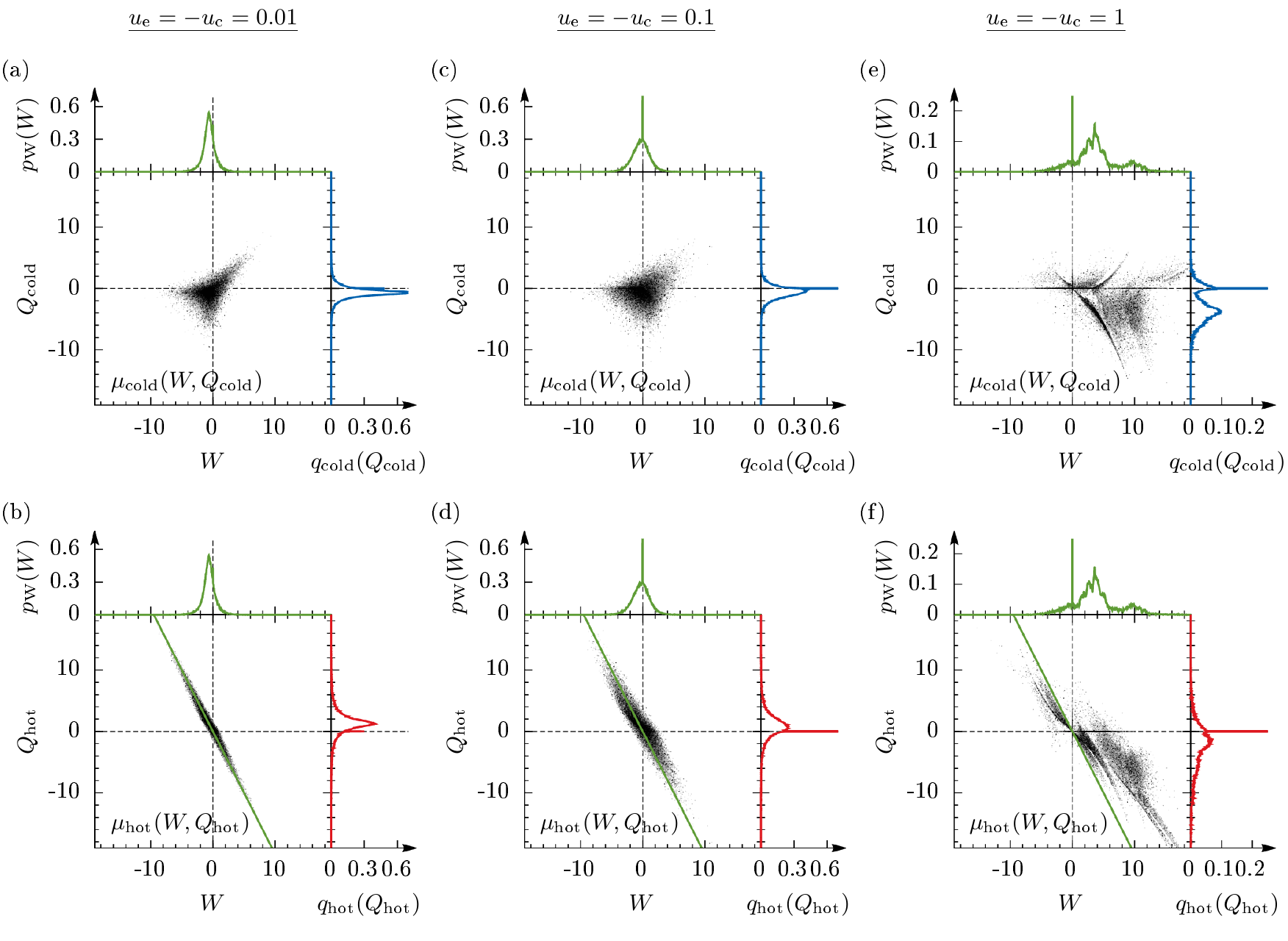}
  \caption{Distribution of work $W$ and heat $Q_\text{cold}$ and $Q_\text{hot}$ exchanged with the cold and the hot bath respectively during one cycle for different piston speeds $u_\text{e}=-u_\text{c}$. In each subplot the center plot displays the joint probability $\mu_\text{cold}(W,Q_\text{cold})$ or $\mu_\text{hot}(W,Q_\text{hot})$ respectively. The marginal work distribution $p_\text{W}(W)$ is always shown on the top whereas the  marginal distributions $q_\text{cold}(Q_\text{cold})$ and $q_\text{hot}(Q_\text{hot})$ are given at the right. Every black dot represents one of $10^5$ simulated cycles. Parameter values are $\beta_\text{cold}^{-1}= 1$, $\beta_\text{hot}^{-1}= 2$, and $u_\text{e}=-u_\text{c}=0.01$ (a,b), $u_\text{e}=-u_\text{c}=0.1$ (c,d), and $u_\text{e}=-u_\text{c}=1$ (e,f). In all cases the length of the cylinder changes in the isothermal stroke at the higher temperature from $L_2=1$ to $L_3=2$. The solid line in subfigures b,d, and f indicates the quasi-static limit.}
  \label{fig:fast_piston}
\end{figure*}

We now come to the general case of arbitrary  $u_\text{e}$ and $u_\text{c}$, where the TPSS operation is of particular interest. By definition, for the TPSS operation the relation $p(x_1,v_1) = p(x_4,v_4)$ must hold. It seems to be difficult to determine these distributions analytically. We therefore rely on numerical results to characterize the TPSS. To this end we start simulations in thermal equilibrium and let them run, until the distributions $p_\text{W}(W)$, $q_\text{hot}(Q_\text{hot})$ and $q_\text{cold}(Q_\text{cold})$ no longer change. All numerical data discussed in this subsection were obtained after such a transient phase. Results for the distributions of work and heat are shown in Fig.~\ref{fig:fast_piston} for three different speeds of the piston.

In contrast to the  quasi-static limit, for finite $u_\text{e}$ and $u_\text{c}$, the relations $W_1 = \Delta F_\text{cold}$ and  $W_3 = \Delta F_\text{hot}$ no longer hold. Yet, if the piston speed is not too large the work is still transferred in many independent collisions the number of which decreases with increasing $u_\text{c}$ and $u_\text{e}$. By virtue of the central limit theorem one would hence expect that for small piston speeds the work variables $W_1$ and $W_3$ become Gaussian distributed with their width increasing with increasing piston speed. This broadening of the distributions $p_\text{W}(W)$, $q_\text{cold}(Q_\text{cold})$, and $q_\text{hot}(Q_\text{hot})$ can be clearly seen by comparing Fig.\ \ref{fig:quasi_stat} with Figs.~\ref{fig:fast_piston}(a-d). When the piston speed increases further the probability for the particle to neither reach piston nor bottom during an entire cycle becomes noticeable. These realizations do not transfer work or heat and give rise to $\delta$-peaks in the distributions of $W$,  $Q_\text{cold}$ and $Q_\text{hot}$, cf. Fig.~\ref{fig:fast_piston}(c-f). Moreover, typical particle trajectories now involve a few collisions only and the distributions of work and heat exhibit a much more spiky structure than those from the case of slow driving, see  Figs.\ \ref{fig:fast_piston} (e) and (f).  Nevertheless we again find exponential tails for the marginal heat distributions shown in Figs.\ \ref{fig:fast_piston} similar to the results of \citep{VanZon2004} for a Brownian particle in a moving parabolic potential. We can no longer test our results against the fluctuation theorem for heat engines \citep{Sinitsyn2011} since the latter is valid only for cycles starting in equilibrium.

\begin{figure*}
  \includegraphics{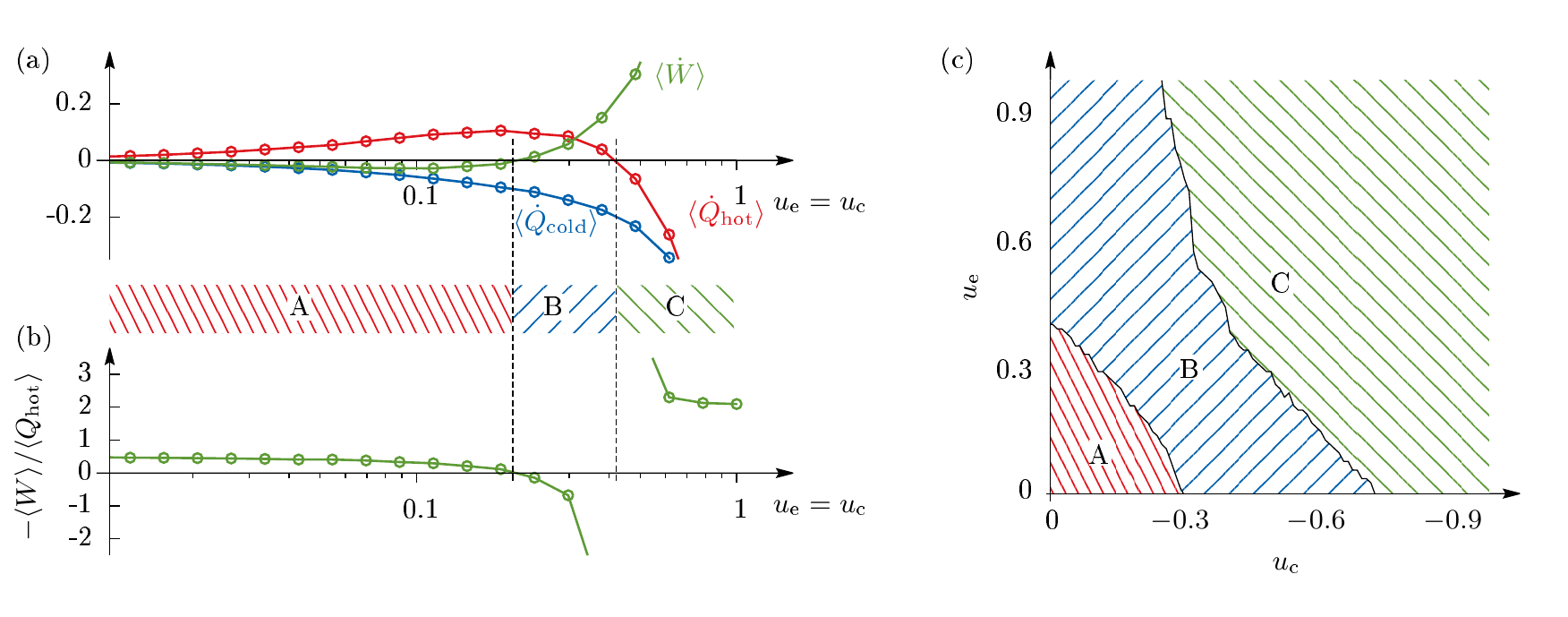}
  \caption{ (a): Mean total power $\langle\dot W\rangle$ and mean heat fluxes $\langle \dot 
   Q_\text{cold,hot} \rangle$ during the hot and the cold isothermal stroke respectively in dependence of the piston speed $u$.  The simulations are conducted for $2\cdot 10^5$ cycles each, with $\beta_\text{cold}^{-1}=1$ and $\beta_\text{hot}^{-1}=2$, and the position of the piston changing from $L_2=1$ to $L_3=2$ during the isothermal expansion at $\beta_\text{hot}^{-1}$.
   (b): Average efficiency obtained from the same simulations.
There are three regimes of operation: $\langle\dot W\rangle < 0 $ (\textbf{A}), $\langle\dot W\rangle > 0$ and $\langle\dot Q_\text{hot}\rangle > 0$ (\textbf{B}), $\langle\dot Q_\text{hot}\rangle < 0$ (\textbf{C}).
(c): Regimes of operations as obtained from simulations of $10^4$ combinations of piston speeds $u_\text{c}$ and $u_\text{e}$.}
  \label{fig:regimes}
\end{figure*}

Based on the average power $\langle\dot W\rangle$ delivered by the cycle and the average heat fluxes $\langle\dot Q_\text{cold}\rangle$ and  $\langle\dot Q_\text{hot}\rangle$  we may distinguish three different regimes of operation as shown in Fig.\ \ref{fig:regimes}. The mean power $\langle\dot W\rangle$ is defined as the total average work per cycle  $\langle W\rangle$ divided by the duration of one cycle and the heat fluxes  $\langle\dot Q_\text{cold}\rangle$ and  $\langle\dot Q_\text{hot}\rangle$ as the average total heat transfers per cycle divided by the duration. In (A) the piston moves slowly and the engine produces power. The efficiency 
$\eta = -\langle W \rangle / \langle Q_\text{hot} \rangle $ starts from the Carnot value $\eta_\text{C}$ in the quasi-static limit and decreases monotonically down to zero at the border of this regime where $\langle\dot W\rangle = 0$. In the intermediate range (B) of piston speeds, the engine consumes work and heat is transferred from the hot to the cold bath. Here $\eta $ is negative. If the piston moves still faster, (C), work is consumed and both reservoirs are heated up. In this regime, $\eta$ is formally larger than one, but the device is clearly useless as a heat engine.

\begin{figure}
  \includegraphics{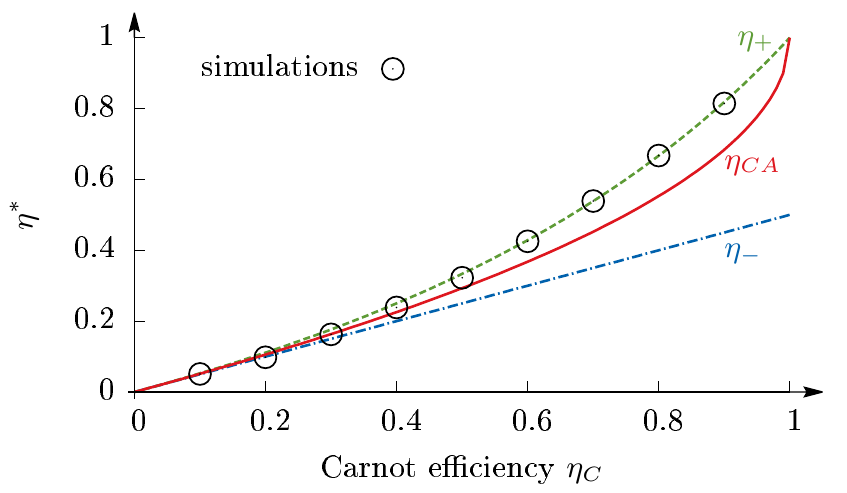}
  \caption{Comparison of the efficiency at maximum power $\eta^*$ (red circles) for different temperature combinations with
    the Curzon-Ahlborn-efficiency  $\eta_{CA}$ (solid line) and the lower and upper bounds $\eta_-$ (dashed-dotted line) and $\eta_+$ (dashed line) obtained in  \citep{ScSe,EsKaLiBr}. The power was optimized using simulations of $10^5$ cycles each for which the cylinder length again changes from $L_2=1$ to $L_3=2$ during the isothermal expansion at the higher temperature.}
  \label{fig:EMP}
\end{figure}

Fig.\ \ref{fig:regimes} also shows, that the power delivered by the engine has a maximum at an intermediate value of $u_\text{e}=-u_\text{c}$. We have determined the {\em efficiency at maximum power} (EMP) $\eta^*$ of our model engine for nine different combinations of $\beta_\text{hot}$ and $\beta_\text{cold}$. For each combination we varied $u_\text{e}$ and $-u_\text{c}$ {\em independently} between $0.01$ and $0.35$ to find the point of maximum power.
The results are shown in Fig.~\ref{fig:EMP} together with the Curzon-Ahlborn efficiency 
$ \eta_{CA} = 1-\sqrt{1-\eta_C}$ \citep{Curzon1975a} and the bounds $  \eta_- = \eta_C/2$ and 
$\eta_+  = \eta_C/(2-\eta_C)$ derived in \citet{ScSe,EsKaLiBr}. Our findings are consistent with these bounds and almost saturate the upper one. The deviation from the Curzon-Ahlborn efficiency is not surprising since these authors assumed a well defined temperature of the working medium whereas our simulations show that the velocity distribution $p(v)$ at maximum power is markedly different from a Maxwell distribution. We also note that the numerical results for the EMP reported in \cite{IzOk1} are somewhat larger than ours thereby violating the upper bound $\eta_+$. This might be due to the fact that in this study the optimization of the power was restricted to $u_\text{e}=-u_\text{c}$, a relation that our optimized values for $u_\text{e}$ and $u_\text{c}$ do not fulfill.

\section{Summary}
\label{sec:concl}

In the present paper we gave a detailed analysis of the stochastic energetics of isothermal compressions and expansions of a classical ideal gas consisting of a single particle. In our model the volume of a one-dimensional cylinder is changed by a piston moving with constant speed $u$. The enclosed particle performs work by elastic collisions with this piston and exchanges heat in inelastic collisions with the bottom of the cylinder. The kinetic energy of the particle is of the order of the average thermal energy per degree of freedom. Accordingly, both work and heat are strongly fluctuating variables, and our focus was on the determination of their probability distributions. The piston speed is arbitrary, and in addition to quasi-static volume changes we were in particular interested in situations where $u$ is comparable to the typical velocity of the particle. 

We first analyzed isothermal compressions and expansions. Introducing a logarithmic time scale we were able to disentangle correlations between successive collisions of the particle which had hampered analytical progress so far. Using elements of renewal theory we then calculated the characteristic function of the work distribution analytically. The distribution itself was determined using a numerical implementation of the inverse Fourier transform. The results agree very well with molecular dynamics simulations and fulfill the Jarzynski equality as well as the Crooks relation. In the quasi-static limit we recover the result that the work ceases to fluctuate and coincides with the difference in free energy between the final and the initial state. For increasing piston speed we first find a broadening of this $\delta$-distribution into a Gaussian before at even larger values of $u$ the typical number of collisions becomes of order one and a multi-modal shape of the work distribution emerges.

We then combined these findings with the results of Lua and Grosberg for adiabatic volume changes to analyze a Carnot-like cycle involving two isothermal strokes at different temperatures and two adiabatic strokes connecting them. For the quasi-static limit we derived analytic results for the joint distribution of the total work and the heat exchanged with the hot reservoir, as well as for the characteristic function of the heat exchanged with the cold bath. Again the results are in very good agreement with molecular dynamics simulations. Somewhat surprisingly strong fluctuations in work and heat remain even in the quasi-static limit: only when considering a large ensemble of independent particles the results of macroscopic thermodynamics are recovered. 

Of special interest are again situations beyond the quasi-static regime. Unfortunately, it seems very hard to determine the invariant density for particle position and velocity of the time periodic steady state describing the stationary operation of the cycle. Therefore we had to rely exclusively on numerical simulations to explore the non-equilibrium performance of the cycle. For increasing piston speed the strong correlations between work and heat observed in the quasi-static limit get weaker. The work delivered per time increases and at intermediate values of the piston speed there is a maximum of the power output. We determined the efficiency at maximum power for several values of the parameters and showed that they are consistent with recently derived bounds. When the piston speed increases further a complicated, spiky shape of the probability distributions emerges. At the same time the efficiency deteriorates until finally the cycle ceases to work as a heat engine altogether: on average it then consumes work and heats up both reservoirs. 

It should be interesting to extend our investigations to the case where the position of the piston is not deterministically prescribed but is itself a fluctuating variable. It would then be possible to analyze the influence of intermingled forward and backwards steps which are characteristic for nanoscopic machines.

\acknowledgments We would like to thank Chris Van den Broeck and David Lacoste for stimulating discussions.

\raggedright

\bibliographystyle{apsrev}

\end{document}